\documentclass[twocolumn,showpacs,amsmath,amssymb,pra]{revtex4}
\usepackage{graphicx}
\usepackage{dcolumn}
\usepackage{bm}

\usepackage[all]{xypic}
\usepackage{verbatim}
\usepackage{amsthm}
\usepackage{graphicx}
\usepackage{color}

\newcommand{\ket}[1]{|#1\rangle}

\newcommand{\ketbra}[2]{|#1\rangle \langle #2 |}
\newcommand{\smnode}{{}\drop\frm<4pt>{o}}

\newcommand{\link}[1]{\ar@{-}[#1]}

\newcommand{\gaussian}[1]{	
	\crv{(-14.2,0.185)!C+<#1,0mm>&(-4.8,2.26)!C+<#1,0mm>&(-0.1,5.41)!C+<#1,0mm>&(4.6,10.11)!C+<#1,0mm>&(9.3,14.71)!C+<#1,0mm>&(14,16.67)!C+<#1,0mm>
	&(18.7,14.71)!C+<#1,0mm>&(23.4,10.11)!C+<#1,0mm>&(28.1,5.41)!C+<#1,0mm>&(32.8,2.26)!C+<#1,0mm>&(42.2,0.185)!C+<#1,0mm>}
	}

\begin{document}

\title{One-way quantum computing in optical lattices with many atom addressing}

\author{Timothy P. Friesen}
\author{David L. Feder}
\affiliation{Department of Physics and Astronomy and Institute for Quantum
Information Science, University of Calgary, Calgary, Alberta, Canada T2N 1N4}

\date{\today}

\begin{abstract}
One of the fundamental conditions for one-way quantum computation (1WQC) is
the ability to make sequential measurements on isolated qubits that comprise
the highly entangled resource for 1WQC, the cluster state. This has been a
significant impediment in the implementation of 1WQC with ultracold atoms
confined in optical lattices, because the width of the measuring lasers
is generally much greater than the atomic (qubit) spacing. We demonstrate that
deterministic 1WQC is nevertheless possible, with a polynomial increase in the
number of operations, as long as the center of the beams can be positioned
with high accuracy. Extending the number of cluster atoms, the scheme is also
able to compensate for accidental measurements of an arbitrary number of
nearby qubits.
\end{abstract}


\maketitle 

\section{Introduction}
\label{introduction}

In one-way quantum computing (1WQC), first proposed by Raussendorf and
Briegel~\cite{RB01}, the computation proceeds entirely by performing a
sequence of single qubit measurements on a special many qubit entangled state,
known as a cluster state. This approach is wholly equivalent to the quantum
circuit model, but provides two distinct practical advantages. First, all of
the required entanglement is generated at the outset in only two steps,
involving all of the particles simultaneously. Second, no entangling gates
are required between arbitrary (and potentially distant) qubits during the
computation. The 1WQC approach therefore lends itself to any physical
architecture where one can induce genuine multipartite entanglement and
perform single-qubit measurements. To date, 1WQC has been demonstrated in
linear optical systems by the implementation of Grover's search algorithm for
up to four qubits~\cite{Walther05}.

Ultracold atoms confined in optical lattices arguably constitute the most
promising test beds for quantum simulation and computation, and are natural
candidates for the implementation of 1WQC. This optimism is due to the 
exceptional control of long-lived atomic internal states, the lattice
parameters, and the interactions between particles. Under the right
conditions, exactly one bosonic atom will occupy each site of the optical
lattice at low temperatures as the gas undergoes a quantum transition into a
Mott insulating phase~\cite{Greiner02}. By adjusting the orientation and
strength of the applied lasers, the Mott phase can be realized for atoms in
effective one-dimensional (1D), 2D, or 3D lattices~\cite{Kohl05}. Because two
internal states can be chosen as computational registers, each atom
corresponds to a qubit. In the Mott phase, these (neutral) qubits are
effectively stationary and non-interacting. The massively entangled cluster
state that is the central resource for 1WQC can then be generated with relative
ease~\cite{BC99,JC00,DD03}: starting with a single large quasi-2D array of 
atoms~\cite{Stock05,Spielman07},
entanglement between neighboring atoms can be done in parallel with
state-dependent collisions~\cite{JB99,Mandel03} or with tunable spin-spin
interactions~\cite{Duan03,Kuklov03}. Allowing these operations to occur for
the right amount of time on qubits initialized in the zero eigenstate of the
Pauli $X$ operator, one can implement (up to local unitaries) the maximally
entangling controlled-phase ($CZ$) gate that generates the cluster state.

While the relative ease with which one may generate large cluster states
containing thousands to millions of physical qubits appears to strongly 
favor ultracold atoms in optical lattices, these systems tend to suffer from
one important shortcoming: the practical difficulty of measuring single
qubits. Most optical lattice experiments to date utilize $^{87}$Rb 
atoms~\cite{Greiner02,Mandel03}, with a resonant absorption at a wavelength of
780~nm. The wavelengths of the lasers forming the optical lattice are
generally chosen to be sufficiently close to this resonance (usually
$\lambda\approx 800$~nm) in order to yield strong confinement while minimizing
spontaneous emission which causes heating and loss of atoms. Assuming
perfectly counter-propagating lasers, the spacing between sites of the optical
lattice is therefore $a\approx 0.4~\mu$m. Yet the additional lasers that would
be used to apply rotations or measurements on selected qubits can generally be
focused to widths on the order of $3-4~\mu$m, which is a few times the
fundamental resolution limit.

Much effort in recent years has been devoted to improving the addressability
of single qubits in optical lattices. An experimental scheme has been realized
wherein atoms occupying every site of an optical lattice are transferred to 
every third site~\cite{Peil03}; in principle this process can be repeated
until the desired separation is reached. A CO$_2$ laser generates an optical
lattice with the long spacing $a=5.3~\mu$m~\cite{Scheunemann00}; while a Mott
transition in this lattice is difficult to achieve, unit-filling in a finite
region may be reached by physically moving atoms through state-dependent laser
manipulation~\cite{weiss04}. This kind of state-dependent transport can also 
be used to move a small number of `marker' atoms through the lattice, 
generating entanglement with the cluster qubits~\cite{Calarco04,Kay04,Kay06};
measurements of the (well-separated) marker atoms can then be unambiguously
made. A recent theoretical proposal makes use of the energy shifts induced by
additional lasers to ensure that only one of the atoms is resonant with the
measuring laser~\cite{zhang06}. Another uses interference patterns generated
by multiple additional lasers oriented at different angles to localize
atoms~\cite{Joo06}; unfortunately, it is generally not feasible experimentally
to generate beams from very many directions or to have the attendant optical
access. A very recent proposal is to pump atoms in the vicinity of the target
into internal states that are decoupled from the subsequent 
manipulations~\cite{Cho07}.

In the present work we show that high-fidelity 1WQC does not in fact require
single-qubit measurements, and in principle is experimentally feasible with
lasers that simultaneously impinge on a large number of atoms from the same
direction. The spatial variation of the Gaussian measuring beam is the crucial
ingredient making this possible. In general, projective measurements on
multiple qubits generally yield mixed states because the lack of spatial
resolution prevents knowing which qubits were projected into the computational
basis states $|0\rangle$ and $|1\rangle$. In order to reduce the uncertainty,
one could simply re-measure different subsets of the qubits. In this scenario
one would expect to require $N-1$ additional measurements for each initial
projection of $N$ qubits to reduce the uncertainty to zero and
yield a pure state. An even simpler approach would be to arrange that the
states of all qubits to be irradiated (except the one of interest) are in a
simple fiducial state, such as one of the computational basis vectors, so
that the ambiguity in the result of the multi-qubit measurement is always
zero. In this second scenario, one would expect $N-1$ additional unitary
operations in advance of each measurement. As discussed in detail below, it is
always possible in principle to implement the second approach, ensuring
high-fidelity 1WQC even for large numbers of simultaneously measured qubits.

The manuscript is organized as follows. The basic ideas of 1WQC and the
fundamentals of implementing rotations and measurements for atoms in optical
lattices are reviewed in Section~\ref{review}. The details of how to initialize
atoms in the states required to form one-dimensional cluster states with many
atom rotations are described in Section~\ref{carving}, and the protocol for
performing (non-universal) 1WQC with multiple measurements of atoms on a line
is discussed in Section~\ref{protocol}. The full two-dimensional protocol is
described in Section~\ref{2dcluster}. The results are summarized in
Section~\ref{summary}, with comments on how the work could be extended.

\section{1WQC and Single Atom Operations}
\label{review}

\subsection{One-way Quantum Computing}
\label{1WQC}

Starting with a periodic two dimensional array of $N$ qubits, a cluster state
can be prepared by first initializing all $N$ qubits to the $+1$ eigenstate of
the Pauli $X$ operator $\ket{+}=(\ket{0}+\ket{1})/\sqrt{2}$, followed by
the application of controlled-phase $CZ={\rm diag}(1,1,1,-1)$ gates between
all nearest neighbors. Computational measurements, made in the eigenbasis of
the Pauli $Z$ operator, are then performed to remove qubits from the initial
cluster state and to enable the construction of a computation-specific cluster
state (hereafter denoted a computational cluster state) with the desired
connections. In the simplest (though not most computationally efficient)
approach, the resulting graph state is a spatial representation of a quantum
circuit: horizontal chains of entangled physical qubits encode a single
computational qubit, and vertical links represent two-qubit entangling
operations.

Actual computation proceeds via quantum teleportation by sequentially measuring
the state of each qubit in a basis defined by the states 
$|\pm_{\xi}\rangle=\left(|0\rangle\pm e^{-i\xi}|1\rangle\right)/\sqrt{2}.$
Alternatively, the state of each qubit is rotated by an angle $\xi$ around the 
$Z$ axis, followed by an $X$-basis measurement. The $X$ measurement can be 
effected by first applying a Hadamard $H$ operation which rotates from the $X$ 
basis to the $Z$ basis, followed by a computational basis measurement.
In a 1D cluster, measuring the first qubit initially in the state
$|\psi\rangle$ teleports the modified state $X^mHR_z(\xi)|\psi\rangle$ to its
nearest neighbor, where we define $R_{\sigma}(\xi)=e^{-i\frac{\xi}{2}\Sigma}$
with $\sigma=x,y,z$ and $\Sigma=X,Y,Z$. The
measurement outcomes $m=0$ and $1$ correspond to projection into the
computational states $|0\rangle$ and $|1\rangle$, respectively. 
Universal single-qubit operations can be decomposed into three successive 
rotations around orthogonal axes (the Euler angles), accomplished in the one-way 
model by three successive measurements:
\begin{eqnarray}
|\psi'\rangle&=&X^{m_3}HR_z(\theta_3)X^{m_2}HR_z(\theta_2)X^{m_1}HR_z(\theta_1)
|\psi\rangle\nonumber \\
&=&X^{m_3}Z^{m_2}X^{m_1}HR_z(\theta_3')R_x(\theta_2')R_z(\theta_1)|\psi\rangle,
\label{primitive}
\end{eqnarray}
where $\theta_2'=(-1)^{m_1}\theta_2$ and $\theta_3'=(-1)^{m_2}\theta_3$. Note
that $R_x(\theta)\equiv\exp(-i\theta X)$.
The Hadamard and the byproduct unitaries (the $X$ and $Z$ in
Eq.~(\ref{primitive}) above)
arising from measurement outcomes $m_i=1$ can wait to be applied only at
the end of the gate teleportation. Crucially the angles $\theta_i'$ for
subsequent measurements must be adapted based on the result of previous
measurements for the computation to be deterministic, a property known as
`feed-forward.' Together with the vertical links in the
computation-specific cluster state, representing $CZ$ gates between
computational qubits, measurements in 1WCQ generate a set of gates that are
universal for quantum computation.

\subsection{Single Atom Rotations}

Rotations on atoms are effected either by a radio-frequency pulse resonant on
the $\ket{0} \leftrightarrow \ket{1}$ transition or by a Raman transition
between these levels using two lasers. In either case, the electromagnetic
field associated with the radiation gives rise to the Hamiltonian in the
dipole and rotating wave approximations~\cite{Walls94}:
\begin{equation}
H_I = \frac{\hbar\Omega}{2}\left(\ketbra{1}{0}e^{i\varphi}
+\ketbra{0}{1}e^{-i\varphi}\right),
\end{equation}
where $\Omega$ and $\varphi$ are the Rabi frequency and laser phase,
respectively. The time evolution operator $U=e^{-iHt/\hbar}$ is then
\begin{equation}
U(\Omega t,\varphi) = \begin{pmatrix}
\cos(\frac{\Omega t}{2}) & -ie^{-i\varphi}\sin(\frac{\Omega t}{2}) \\
-ie^{i\varphi}\sin(\frac{\Omega t}{2}) & \cos(\frac{\Omega t}{2}) 
\end{pmatrix},
\end{equation}
where $U(\Omega t,0)=R_x(\Omega t)$ and $U(\Omega t,\pi/2)=R_y(\Omega t)$.

Using the identity $H=-iR_x(\pi)R_y(\pi/2)$ we can write
$HR_z(\theta)=R_x(\theta)H=R_x(\theta+\pi)R_y(\pi/2)$ neglecting the overall
phase. Because an $R_y(\pi/2)$
is always present independent of the angle $\theta$, one can simultaneously
apply $R_y(\pi/2)$ with a wide laser beam to all the atoms in the
computational cluster state before making any measurements. In this way only 
$R_x$ rotations need to be applied to the qubits before measuring in the
computational basis: the result of the teleportation is the same as if the
rotation had been applied after measurement of the previous qubit.

The focused rotation laser is unfortunately not as narrow as the lattice
spacing, and will therefore irradiate atoms near the qubit of interest.
The beam intensity has a Gaussian distribution $I=I_0\exp(-2x^2/r^2)$ where
$x$ is the distance from the center and $r$ is the beam radius at which 
$I=I_0e^{-2}$. The Rabi frequency $\Omega$ is proportional to the electric
field and therefore the square root of the intensity 
$\Omega(x)=\Omega_0\exp(-x^2/r^2)$. All neighboring atoms are influenced by
the field for the same pulse length $t$, so atoms further from the beam center
will undergo a rotation by an exponentially decreasing angle.

\subsection{Single Atom Measurements}\label{measec}

Measuring the internal state of an atom is done by fluorescence, a technique
used for quantum jump detection~\cite{BZ88,PK98,GM93}. Atomic measurements of
this type have been experimentally demonstrated with ions (where the technique
is also known as electron shelving) and work with high
 probability~\cite{LB03}. In a
fluorescence measurement a pulse is applied between the computational state 
$\ket{0}$ and an unstable auxiliary state $\ket{2}$ which rapidly decays
spontaneously back to $\ket{0}$ (the decay pathway $\ket{2}\rightarrow\ket{1}$
is forbidden). The atom will repeatedly transition between $\ket{0}$ and
$\ket{2}$ producing many fluorescent photons as long as the measurement pulse
is active, while an atom in $\ket{1}$ will not be affected by the measurement
beam. A strong measurement corresponds to the projection of the atom's internal
state wavefunction into one of the computational basis states with high
certainty. 

While the measurement pulse is active, the Hamiltonian for the system in the
dipole and rotating wave approximations is given by: 
\begin{equation}
H=\frac{\hbar \Omega}{2}(\ketbra{0}{2}+\ketbra{2}{0}).
\end{equation}
where $\Omega$ is the $\ket{0}\leftrightarrow\ket{2}$ Rabi frequency. The time
evolution of the atomic density matrix is described by the master 
equation~\cite{Walls94}
\begin{equation} \dot \rho = -\frac{i}{2}[H,\rho] 
+\frac{\gamma}{2}(2\ketbra{0}{2}\rho\ketbra{2}{0}-\ketbra{2}{2}\rho
-\rho\ketbra{2}{2}) 
\end{equation}
where $\gamma\gg\Omega$ is the rapid decay rate from excited level $\ket{2}$.
For times long compared to $1/\gamma$ one can effectively eliminate the
population in $\ket{2}$. Neglecting terms proportional to $(\Omega/\gamma)^2$
one obtains
\begin{eqnarray}
\dot\rho_{00}&=& 0; \\
\dot\rho_{11}&=& 0; \\
\dot\rho_{01}&=& -\frac{\Omega^2}{2\gamma}\rho_{01}.
\end{eqnarray}
These equations describe a two-level system with a decay of the $\rho_{01}$
coherence a rate $\frac{\Omega^2}{2\gamma}t$. A successful measurement
requires the vanishing of the off-diagonal elements (coherences) of the
density matrix, $\frac{\Omega^2}{2\gamma}t\gg 1$, which together with
$\Omega/\gamma\ll 1$ requires $t\gg\Omega^{-1}$. The probability of a
measurement occurring is therefore given by
\begin{equation} 
p=1-e^{-\frac{\Omega^2}{2\gamma}t}
\label{prob}
\end{equation}
which approaches unity exponentially for high intensities, strong coupling,
rapid decay times, and long measurements.

As was the case for rotations, the spatial profile of the measuring laser
ensures that atoms near the qubit of interest will have a reasonably high
(though exponentially decreasing) probability of being measured. The 
consequences would appear to be disastrous for 1WQC, because unwanted 
projections of qubits in the computational cluster will completely destroy
the state of the logical qubit. Furthermore, an inadvertent measurement will
be nearly impossible to detect. The fluorescence is proportional to the local
beam intensity, so that atoms projected by the low-intensity wings of the
Gaussian will emit comparatively few photons. The solution to this apparent
impasse will be discussed in Section~\ref{protocol}.

\section{Cluster state creation}
\label{carving}

The first hurdle facing practical 1WQC in an optical lattice system is the
formation of the computational cluster state. The usual approach taken is
to first create a standard cluster state and then to selectively remove
unwanted physical qubits through $Z$ measurements. This method is not 
feasible for wide measuring beams, however. There will always be some
probability of inadvertently (and unknowingly) measuring the states of nearby
qubits, thereby removing atoms that should form part of the cluster. A more
suitable construction is to first rotate all computational cluster qubits to
$\ket{+}$ and all unwanted qubits to $\ket{0}$, and only then perform the
controlled phase gates between all neighboring qubits. Qubits in $\ket{0}$
will be left untouched by the $CZ$ operations and therefore will not be
connected to neighboring qubits.

It is possible to specify the states of individual qubits in an optical
lattice even for wide rotation lasers that irradiate multiple atoms. One
requires only that the {\it center} of the beam can be positioned with high
accuracy. Consider an array of $N$ qubits on which one performs $N$ rotation
pulses, each centered on a different qubit, with angles $\theta_n$ about the
same axis, where $n=1,2,...,N$. The beams have a Gaussian intensity profile
and therefore will rotate each qubit $m$ by an angle 
$\phi_m=\theta_ne^{-x_{mn}^2/r^2}$, where $x_{mn}=|x_n-x_m|$ is the distance
from the central qubit $n$. This will lead to a linear system of equations with 
$N$ unknowns (the $\theta_n$) that yield the desired phases (the $\phi_n$) on
each qubit. In general, the linear system takes the form

\begin{equation}
A\vec{\theta}=\vec{\phi}
\label{Axv}
\end{equation}
where $A$ is of the form
\begin{equation}
 A = \begin{pmatrix}
      1 & a_{12} & a_{13} & a_{14} & ... & .\\
      a_{21} & 1 & a_{23} & ... & &.\\
      a_{31} & a_{32} & 1 & ... & &.\\
      . & & & . & &\\
      . & & & & . & & \\
      . & .&. & & & . &  \\
\end{pmatrix},
\end{equation}
\begin{equation}
a_{nm}=e^{-x_{nm}^2/r^2}=a_{mn},
\end{equation}
\begin{equation}
\vec{\theta} = \begin{pmatrix} 
\theta_1 \\
\theta_2 \\
\theta_3 \\
. \\
.\\
.\\
\end{pmatrix},
\qquad
\vec{\phi} = \begin{pmatrix}
\phi_1 \\
\phi_2 \\
\phi_3 \\
. \\
.\\
.\\
\end{pmatrix}.
\end{equation}

In order for a solution to exist one requires
\begin{equation}
\displaystyle\det(A)=\sum_{k}b_{k}e^{-k/r^2}\neq 0,
\end{equation}
where both the $k$ (sums of distances squared in the $x$ and $y$ directions of
the cluster) and the $b_k$ (combinatoric coefficients) are integers. Suppose
one assumes that $\det(A)=0$. This is equivalent to stating that $e^{-1/r^2}$
is a root of a polynomial with integer coefficients. For rational $-1/r^2$,
however, $e^{-1/r^2}$ is transcendental which by definition cannot be the 
root of a polynomial with integer coefficients~\cite{B90}. It follows that
$\det(A)$ is non-zero and a solution exists for the linear 
equations~(\ref{Axv}). Therefore, with one beam centered on each of the $N$
atoms it is possible to precisely control the rotation applied to each
individual atom. 

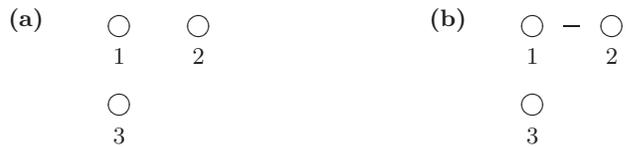
\begin{figure}[t]
\begin{center}
$\begin{array}{c@{\hspace{1in}}c}
\multicolumn{1}{l}{\mbox{\bf (a)}} &
\multicolumn{1}{l}{\mbox{\bf (b)}} \\ [-0.53cm]
\xymatrix @=6pt @*=<24pt,24pt> {
          %1          2       
    \ & \smnode \save[]+<0mm,-4mm>*\txt{1} \restore & \smnode \save[]+<0mm,-4mm>*\txt{2} \restore  \\ %1
    \ & \smnode \save[]+<0mm,-4mm>*\txt{3} \restore &  \\ %2
  } &
	\xymatrix @=6pt @*=<24pt,24pt> {
          %1                           2       
    \ & \smnode \save[]+<0mm,-4mm>*\txt{1} \restore \link{r} & \smnode \save[]+<0mm,-4mm>*\txt{2} \restore  \\ %1
    \ & \smnode \save[]+<0mm,-4mm>*\txt{3} \restore &  \\ %2
  } \\ [0.4cm]
\end{array}$
\end{center}
\caption{Building a two-qubit cluster state with three qubits. The initial
state (a) has all three qubits in state $\ket{+}$; after rotations qubit 3 is
in state $\ket{0}$, so that entanglement leaves the desired final state (b).}
\label{3qbit}
\end{figure}

Consider a simple example in order to demonstrate the procedure. Suppose there
are three qubits oriented as shown in Fig.~\ref{3qbit}, and one wishes to
generate entanglement between the qubits 1 and 2 but leave qubit 3
disentangled, Fig~\ref{3qbit}(b). First all three qubits are initialized to
$\ket{+}$ with a very wide beam (which assumes negligible spatial variation
in the intensity). Three $R_y(\theta_i)$ pulses are then applied, each one
centered on a different qubit, such that qubits 1 and 2 remain in $\ket{+}$
while qubit 3 returns to $\ket{0}$. The linear system is:
\begin{eqnarray}
1:\theta_1 + e^{-1/r^2}\theta_2 + e^{-1/r^2}\theta_3&=&0;\\
2:e^{-1/r^2}\theta_1 + \theta_2 + e^{-2/r^2}\theta_3&=&0;\\
3:e^{-1/r^2}\theta_1 + e^{-2/r^2}\theta_2 + \theta_3&=&-\pi/2,
\end{eqnarray}
which has the solution
\begin{eqnarray}
\theta_1&=&\frac{\pi}{2}\frac{e^{-1/r^2}}{1-e^{-2/r^2}};\\
\theta_2&=&0;\\
\theta_3&=&-\frac{\pi}{2}\frac{1}{1-e^{-2/r^2}}.
\end{eqnarray}
In this particular case only two rotations are actually needed. Note that the
solution depends on the beam radius $r$ only in terms of an (experimentally)
adjustable parameter, as long as $r$ remains finite. For a larger number of
qubits and a wide beam, each rotation will affect many qubits across the
lattice, but as long as there are as many distinct operations as qubits, each
applied rotation may be chosen such that the overall phase on each qubit is
exactly as desired.

\section{1D Cluster Computation}
\label{protocol}

\subsection{Offset Measurements}

\begin{figure}[t]
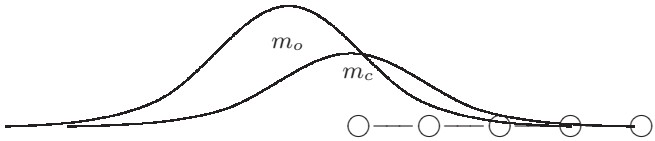

%\begin{center}
\[
\xy
(-23.6,0)!C+<-18.8mm,0mm>;(51.6,0)!C+<-18.8mm,0mm>**\gaussian{-18.8mm};
(-23.6,0)!C+<-9.4mm,0mm>;(51.6,0)!C+<-9.4mm,0mm>**\crv{(-14.2,0.112)!C+<-9.4mm,0mm>&(-4.8,1.37)!C+<-9.4mm,0mm>
&(-0.1,3.28)!C+<-9.4mm,0mm>&(4.6,6.13)!C+<-9.4mm,0mm>&(9.3,8.92)!C+<-9.4mm,0mm>&(14,10.11)!C+<-9.4mm,0mm>&(18.7,8.92)!C+<-9.4mm,0mm>
&(23.4,6.13)!C+<-9.4mm,0mm>&(28.1,3.28)!C+<-9.4mm,0mm>&(32.8,1.37)!C+<-9.4mm,0mm>&(42.2,0.112)!C+<-9.4mm,0mm>};
(-3.8,11)*\txt<2pc>{$m_o$};
(5.6,7)*\txt<2pc>{$m_c$};
(5.6,0)*\frm<4pt>{o};
(8.6,0);(12,0)**\crv{};
(15,0)*\frm<4pt>{o};
(18,0);(21.4,0)**\crv{};
(24.4,0)*\frm<4pt>{o};
(27.4,0);(30.8,0)**\crv{};
(33.8,0)*\frm<4pt>{o};
(36.8,0);(40.2,0)**\crv{};
(43.2,0)*\frm<4pt>{o};
(0,-10)*\frm<4pt>{};
\endxy 
\]
\caption{The advantage of offsetting Gaussian pulses. The centered beam $M_c$ 
has the same intensity on the target qubit as a higher-intensity beam $M_o$
whose center is displaced from the target qubit by a single lattice spacing to
the left, but $M_o$ has an exponentially reduced intensity on the first 
neighbor to the right.}
\label{2beams}
%\end{center}
\end{figure}

The simplest way to reduce the probability of inadvertently measuring qubits in
an entangled chain is to offset the center of the measuring beam. Consider the 
very first qubit in a one-dimensional cluster, Fig.~\ref{2beams}. To the left
of this qubit there is empty space and to the right are connected qubits.
We wish to measure the first qubit with high probability while preserving the
states of its neighbors. This can be accomplished by taking advantage
of the Gaussian shape of the laser pulse: shifting the beam to the left while
increasing its maximum intensity reduces its amplitude on the neighboring
qubits without altering its effect on the target atom.

Suppose that one requires intensity $I=I_t$ on the first (target) qubit.
Shifting the beam center $n$ lattice spacings $a$ to the left yields intensity
on the target $I=I_ne^{-2(na)^2/r^2}$, so that the intensity maximum of the
beam must increase exponentially in the offset: $I_n=I_te^{2(na)^2/r^2}$. The
intensity on the first neighbor to the right of the target $I_{\rm f}$ 
nevertheless decreases exponentially in the offset:
$I_{\rm f}=I_te^{2(na)^2/r^2}e^{-2(n+1)^2a^2/r^2}=I_te^{-2(2n+1)a^2/r^2}$. If
the desired probabilities of measuring the target and nearest-neighbor qubits
are $p_{\rm t}$ and $p_{\rm f}$, respectively, then from Eq.~(\ref{prob})
one obtains the required offset
\begin{equation}
n=\frac{r^2}{4a^2}\ln\left[\frac{\ln(1-p_{\rm t})}{\ln(1-p_{\rm f})}\right]
-\frac{1}{2}.
\label{offset}
\end{equation}
For example, if $p_{\rm t}=0.99$ and $p_{\rm f}=0.01$ were desired, then
with $r=4a$ one would require $n\approx 24$. With $r=10a$ the offset jumps to
approximately $153a$. The probability of simultaneously measuring the second
nearest neighbor $p_{\rm s}$ is truly negligible:
\begin{equation}
p_{\rm s}=1-\exp\left[e^{-4a^2/r^2}\frac{\ln^2(1-p_{\rm f})}{\ln(1-p_{\rm t})}
\right].
\end{equation}
For the same choices of $p_{\rm t}$ and $p_{\rm f}$ above, one obtains 
$p_{\rm s}\approx 10^{-5}$.

The discussion above has assumed that the target qubit is located at the
left-most edge of the chain, with its neighbors to the right. In the most
general case of an offset measurement, however, a significant number of
qubits $M$ to the left of the target will also be measured at each stage of
the computation ($M\approx 2n$ where $na$ is the offset distance discussed 
above).  Because all of these were target
qubits during previous steps in the one-way computation, their subsequent
projections into computational basis states during the measurement of the
target qubit will not affect the logical states of any of the cluster
qubits. That said, rotation pulses will have been applied to these qubits by
the wide lasers during the computation, so their measurement outcomes are not
predetermined.

One must be able to clearly distinguish the fluorescence of the target atom
from that of all the qubits to its left when performing a measurement with an
offset beam. The simplest solution is to re-set the
states of all $M$ non-cluster qubits that will be re-measured to the
non-fluorescing state $|1\rangle$. This can be accomplished with the same
method as described in Sec.~\ref{carving} to initialize all of the qubits'
states. The phases $\varphi_i$ accumulated on the $M$ relevant qubits by
previous rotations are known, so that one can choose $\phi_i=-\varphi_i+\pi$
as input to the linear system of equations to find the
$\theta_i$. Note that the application of the $M$ rotations will also build up
an undesired phase on the target qubit, so this qubit needs to be included in
the linear system, with $\phi_{\rm target}=0$. Thus, $M+1$ operations are
required for each target measurement. In addition, the phases of $2n$ qubits
need to be calculated classically at each stage; together with the time for the
classical solution to the $2n+1$-dimensional linear system, each measurement
has a classical overhead that scales like $4n^2$.

\subsection{Measurement Protocol}

As discussed in detail above, with wide beams one is liable to make 
inadvertent measurements of atoms close to the target qubit, though the
probability of this occurring can be made arbitrarily small in principle
by offsetting the center of the measuring beam from the target. If only one of
the measured qubits to the right of the target is in a state outside the $XY$
plane prior to its measurement, the logical qubit will undergo a non-unitary
(and therefore uncorrectable) transformation. The most extreme case is a $Z$
measurement, which unentangles the measured qubit from the rest of the cluster
and destroys the entanglement resource needed for quantum teleportation. Even
if all measured qubits' states were initially in the $XY$ plane, the absence
of feed-forward makes it unlikely that the desired unitary transformation will
be implemented.

There are two cases when feed-forward is not required, however. The first
is if all of the measurement outcomes are zero, in which case there are no
byproduct operators to commute through the applied rotations, and therefore
no compensation of the choice of measurement angle. Unfortunately, the
likelihood of this occurrence decreases exponentially with the number of
measured qubits, and in any case the results of inadvertent measurements
are by definition unknown. The second case corresponds to teleportation of
single qubit Clifford operations, which byproduct operators can
commute through without changing the operation~\cite{browne-2006}. It is 
impossible, however, to effect universal quantum computation using Clifford 
operations alone.

Below we describe in detail a protocol that eliminates errors arising from
inadvertent measurements of near neighbors, with linear overhead in terms of
ancillary physical qubits and operations. Feed-forward is normally accounted
for by applying a corrective rotation depending on the result of the previous
measurement. As discussed in Sec.~\ref{1WQC}, this corrective rotation need
only be applied when that result is one. When $m_1 = 0$ the gate teleportation
is successful even if an inadvertent measurement of a neighbor has occurred.
When $m_1 = 1$ there will be some probability $p$ that an error has occurred,
but the negative consequences of this can be avoided by implementing
feed-forward in an alternative manner. Rather than applying a corrective
rotation on the next measurement, we resign ourselves to rotating in the wrong
direction. Instead, the sequence of measurements is extended by a number of
qubits on which the phase error can hopefully be corrected. We first describe
in detail the protocol to compensate for inadvertent measurements of the
qubit that is the nearest-neighbor of the target; the extension to the case
where an arbitrary number of qubits is inadvertently measured is discussed at
the end.

The goal is to implement three successive Euler rotations $HR_z(\alpha_i)$,
with the $\alpha_i$ unequal angles. As discussed in Sec.~\ref{1WQC}, these
constitute a universal single-qubit unitary on the computational qubits.
After entanglement, $R_y(\pi/2)$ is applied to all of the qubits so that
future rotations can be performed only about the $X$ axis. To effect the
first two Euler rotations, one applies $R_x$ rotations such that the first
and second qubits are chosen to have a total angle of $\alpha_1+\pi$ and
$\alpha_2+\pi$, respectively. This can be accomplished by applying two
rotation pulses in analogy to the cluster-carving protocol described in
Sec.~\ref{carving}, except now only with $R_x$ rather than $R_y$ rotations.
The first pulse rotates the first qubit by an angle $\theta_1$ and the second
pulse rotates the second qubit by $\theta_2$, where the $\theta_i$ can be
chosen by solving the linear system
\begin{equation}
\alpha_1+\pi = \theta_1 + e^{-1/r^2}\theta_2
\end{equation}
\begin{equation}
\alpha_2+\pi = e^{-1/r^2}\theta_1 + \theta_2
\end{equation}
with $r$ the width of the rotation beam. Solving for $\theta_1$ and
$\theta_2$ one obtains
\begin{equation}
\theta_1 = \frac{\alpha_1+\pi-e^{-1/r^2}(\alpha_2+\pi)}{1-e^{-2/r^2}}.
\end{equation}
\begin{equation}
\theta_2 = \frac{\alpha_2+\pi-e^{-1/r^2}\left(\alpha_1+\pi\right)}{1-e^{-2/r^2}};
\end{equation}
These pulses will apply undesired unitaries to the other qubits in the chain,
but these can be compensated for later. The total applied rotations are then
\begin{eqnarray*}
1:& R_x(\alpha_1+\pi)R_y(\pi/2) \\
2:& R_x(\alpha_2+\pi)R_y(\pi/2) \\
3:& R_x(e^{-4/r^2}\theta_1+e^{-1/r^2}\theta_2)R_y(\pi/2) \\
& \vdots
\end{eqnarray*}
i.e.\ total unitaries of $HR_z(\alpha_1)$ and $HR_z(\alpha_2)$ have now been
applied to qubits 1 and 2, respectively.

Measurement of the qubit 1 teleports the logical state
$X^{m_1}HR_z(\alpha_1)\ket{\psi}$ to the second qubit. Taking into account the
finite
probability of an inadvertent measurement of the second qubit, the logical 
state (defined as the state of the qubit immediately to the right of the 
atom just measured, not including as-yet unmeasured qubits further to the 
right or single-qubit unitaries that have resulted from previous rotations)
is the density matrix:
\begin{eqnarray}
\rho_1&=&(1-p)||X^{m_1}HR_z(\alpha_1)\ket{\psi}||\nonumber \\
&+&p||X^{m_2}HR_z(\alpha_2)X^{m_1}HR_z(\alpha_1)\ket{\psi}||,\nonumber \\
&=&(1-p)||X^{m_1}HR_z(\alpha_1)\ket{\psi}||\nonumber \\
&+&p||X^{m_2}Z^{m_1}R_x[(-1)^{m_1}\alpha_2]R_z(\alpha_1)\ket{\psi}||,
\label{rho1}
\end{eqnarray}
where $||\,|\psi\rangle||\equiv|\psi\rangle\langle\psi|$, $p=p_{\rm f}$, and
the identities $R_z(\alpha)X\equiv XR_z(-\alpha)$ and
$R_x(\alpha)=HR_z(\alpha)H$ have been used to derive the final expression.
Note that the fidelity can never be exactly unity during the computation
because there is always the probability $p$ of inadvertently measuring the
adjacent qubit. 

Now the second qubit must be measured. If the second qubit had already been
inadvertently projected, then the outcome of the second measurement will
definitely be $m_2$; otherwise the value of $m_2$ will be determined presently.
If $m_1=0$, one could carry out this measurement by leaving the first qubit in
its $\ket{0}$ state and distinguishing the second qubit's measurement outcome by
an increased or unchanged level of fluorescence compared to measuring the
first qubit alone. In keeping with the approach outlined in the above section,
however, prior to measurement one should rather apply a total rotation of
$\pi$ to the first qubit to rotate its state into $\ket{1}$ (which
produces no fluorescence signal), and a rotation of $0$ to the second qubit to
preserve its state.

Neglecting for the moment the possibility of inadvertently measuring the third
qubit upon measurement of the second qubit, the result would be the pure
logical state $X^{m_2}Z^{m_1}R_x[(-1)^{m_1}]R_z(\alpha_1)\ket{\psi}$. If
$m_1=0$, then the second Euler rotation will have been correctly effected, and
one could simply apply an angle $\alpha_3+\pi$ to qubit 3 to effect the last
Euler rotation, completing the general single-qubit unitary. (Recall that with
the wide beam one also needs to ensure that all previously intentionally
measured qubits are rotated into $\ket{1}$, and that the accumulated phases on
atoms beyond qubit 4 are noted for future reference). Now explicitly including
the possibility of inadvertently measuring the third qubit, after measurement
of qubit 2 the logical state would be
\begin{eqnarray}
\rho_2&=&(1-p)||X^{m_2}R_x(\alpha_2)R_z(\alpha_1)\ket{\psi}||\nonumber \\
&+&p||X^{m_3}HR_z(\alpha_3)X^{m_2}R_x(\alpha_2)R_z(\alpha_1)\ket{\psi}||
\nonumber \\
&=&(1-p)||X^{m_2}R_x(\alpha_2)R_z(\alpha_1)\ket{\psi}||\nonumber \\
&+&p||X^{m_3}Z^{m_2}HR_z[(-1)^{m_2}\alpha_3]R_x(\alpha_2)\nonumber \\
&&\;\times R_z(\alpha_1)\ket{\psi}||.
\end{eqnarray}
Again, if $m_2=0$, then the third Euler angle will have been applied correctly,
and measurement of qubit 3 will complete the unitary. The rotation on the
fourth qubit is arbitrary, so let's set it to $\pi$. The logical state is 
finally 
\begin{eqnarray}
\rho_3&=&(1-p)||X^{m_3}HR_z(\alpha_3)R_x(\alpha_2)R_z(\alpha_1)
\ket{\psi}||\nonumber \\
&+&p||X^{m_4}HX^{m_3}HR_z(\alpha_3)R_x(\alpha_2)R_z(\alpha_1)
\ket{\psi}||\nonumber \\
&=&(1-p)||X^{m_3}HR_z(\alpha_3)R_x(\alpha_2)R_z(\alpha_1)
\ket{\psi}||\nonumber \\
&+&p||X^{m_4}Z^{m_3}R_z(\alpha_3)R_x(\alpha_2)R_z(\alpha_1)
\ket{\psi}||.
\end{eqnarray}
It is important to note that the overall fidelity has not decreased from the
state (\ref{rho1}) after the first measurement. Each successive measurement
effectively purifies the logical state into the same sum of two density
matrices. The right-most cluster qubit is well-defined and so the very last
measurement can be chosen to be unambiguous, yielding unit fidelity at the end
of the full 1WQC.

The scenario above assumes that $m_i=0$ for all $i$, but the probability of
this occurring decreases exponentially in the number of measured qubits.
Thankfully, post-selecting on these outcomes is not necessary with our 
protocol. Consider the case $m_1=1$. Measuring qubit 2 will result in the
incorrect
Euler angle $-\alpha_2$. This scenario is quite different from standard 1WQC,
whereupon measuring $m_1=1$ the opposite angle would be chosen for the
measurement of qubit 2 to immediately correct the feed-forward error. In the
present case qubit 2 may have already been inadvertently measured, and it is
too late to correct the error in this manner. Rather, measurements are
performed on the next pair of qubits in an attempt to rotate the Euler angle
back to the desired value. Rotations of $\pi$ and $-2\alpha_2+\pi$ are applied
to qubits 3 and 4, respectively. Measuring qubits 2 and 3 then yields the
logical state
\begin{eqnarray}
\rho_3&=&(1-p)||X^{m_3}HX^{m_2}ZR_x(-\alpha_2)R_z(\alpha_1)\ket{\psi}||
\nonumber \\
&+&p||X^{m_4}HR_z(-2\alpha_2)X^{m_3}HX^{m_2}ZR_x(-\alpha_2)\nonumber \\
&&\;\times R_z(\alpha_1)\ket{\psi}||\nonumber \\
&=&(1-p)||X^{m_3}Z^{m_2}XHR_x(-\alpha_2)R_z(\alpha_1)\ket{\psi}||
\nonumber \\
&+&p||X^{m_4}Z^{m_3}X^{m_2}ZR_x[(-1)^{m_3}2\alpha_2]R_x(-\alpha_2)\nonumber \\
&&\;\times R_z(\alpha_1)\ket{\psi}||.
\end{eqnarray}
If $m_3=0$, then the logical state becomes
\begin{eqnarray}
\rho_3&=&(1-p)||Z^{m_2}XHR_x(-\alpha_2)R_z(\alpha_1)\ket{\psi}||\nonumber \\
&+&p||X^{m_4+m_2}ZR_x(\alpha_2)R_z(\alpha_1)\ket{\psi}||,
\end{eqnarray}
so that the correct second Euler would be applied upon measurement of qubit 4.
It suffices to rotate qubit 5 into $\alpha_3+\pi$ in advance, and continue as
if no error had occurred. If rather $m_3=1$, then the applied Euler angle 
would be $-3\alpha_2$, and another qubit pair need to be inserted (the second
with angle $4\alpha_2+\pi$), etc. The probability of {\it not} obtaining a
$0$ outcome on the relevant odd-numbered qubit (and therefore not applying the
correct Euler rotation) decreases exponentially in the number of attempts.

Following this example above, if now $m_3=0$ one can choose qubit 5 to have
the phase $(-1)^{m_2}\alpha_3+\pi$, so that measurement of qubit 4 yields the
logical state
\begin{eqnarray}
\rho_4&=&(1-p)||X^{m_4+m_2}ZR_x(\alpha_2)R_z(\alpha_1)\ket{\psi}||\nonumber \\
&+&p||X^{m_5}HR_z[(-1)^{m_2}\alpha_3]X^{m_4+m_2}ZR_x(\alpha_2)\nonumber \\
&&\;\times R_z(\alpha_1)\ket{\psi}||\nonumber \\
&=&(1-p)||X^{m_4+m_2}ZR_x(\alpha_2)R_z(\alpha_1)\ket{\psi}||\nonumber
\\
&+&p||X^{m_5}Z^{m_4+m_2}XHR_z[(-1)^{m_4}\alpha_3]R_x(\alpha_2)\nonumber \\
&&\;\times R_z(\alpha_1)\ket{\psi}||.
\end{eqnarray}
If $m_4=0$, measurement of qubit 5 yields the desired unitary; 
otherwise, additional pairs of qubits need to be appended to qubit 5 to rotate
the Euler angle back to the desired value, as described above.

We now summarize the protocol. Each $HR_z(\alpha_i)$ operation in principle
requires only one measurement on each physical qubit, all prefaced by the
appropriate rotations on nearby qubits. However, if the sum of certain
previous measurement outcomes is odd, one needs to correct for errors due to
possible inadvertent measurements (even if they haven't occurred!) by inserting
additional pairs of measured qubits. Four qubits is the minimum number for a
universal single-qubit unitary, though in practice the number
could be considerably higher, depending on the measurement outcomes.

While this protocol protects against inadvertent measurements on the qubit
immediately to the right of the target, accidental measurement of the 
next-nearest neighbor could be catastrophic for the overall 1WQC. With offset
measurements the probability of this occurring is small, but for a long
computation the number of such events can become appreciable. A protocol
protecting against inadvertent measurements of many neighbors avoids the
requirement of very large laser offsets with wide beams. It could also provide
a way to protect quantum information against re-absorption by distant atoms
of scattered photons emanating from the fluorescing target.

To protect against more long-range inadvertent measurements requires additional
`buffer' qubits rotated by $\pi$ that effect only Clifford gates. Consider the
case where one intends to compensate for inadvertently measuring two neighbors.
As above, all qubits initially undergo $R_y(\pi/2)$ rotations. The first
two qubits undergo $R_x$ rotations by $\alpha_1+\pi$, $\alpha_2+\pi$ as
before. Now the possibility that the third qubit might be inadvertently
projected suggests that it be rotated by $\pi$ prior to any measurements.
Again, the outcome of the first measurement $m_1$ determines if the second
Euler rotation will be implemented correctly. If $m_1=0$, then one can proceed
directly with implementing the third Euler rotation. One cannot simply rotate
qubits 4 and 5 respectively by $\alpha_3+\pi$ and $\pi$, however. The
combination of $\pi$ and $\alpha_3+\pi$ on qubits 3 and 4 eliminates the
Hadamard gate needed for the next Euler rotation to be around the $z$ axis.
One therefore needs qubits 4 through 6 rotated by $\pi$, 
$(-1)^{m_2}\alpha_3+\pi$, and $\pi$, respectively. Again, $m_4$ determines the
success of
the last Euler rotation. If $m_4=0$ then qubit 7 needs to be set to $\pi$
and the fifth qubit measured to effect the single-qubit unitary. Thus, a
minimum of seven qubits are required to protect against two-qubit inadvertent
measurements. If either $m_1=1$ or $m_4=1$, additional qubit pairs must
be added until the Euler angle is rotated back to the desired value, just as
in the case considered in detail above; in the former case, for example,
qubits 4 and 5 would be rotated by $-2\alpha_2+\pi$ and $\pi$, respectively.

It should now be apparent that to protect against inadvertent measurements on
$m$ qubits to the right of the target requires on the order of $3m$ physical
qubits in order to effect a universal single-qubit gate on computational
qubits. Because feed-forward errors occur with probability $1/2$, the
number of qubits needed to implement the correct Euler angle is of order 
$m$. A much larger number of measurements may be required
in practice to correct an Euler angle error, but the probability of continued
failures becomes exponentially smaller in the number of measurements.

Compensating for inadvertent measurements on multiple qubits is {\it not}
equivalent to making actual measurements on multiple qubits. The former
assumes that fluorescence from inadvertently projected qubits is not
observable while the latter assumes that it is. To simultaneously measure
multiple qubits one could in principle make use of the Gaussian profile of the
measuring pulse: the qubit fluorescence signals are proportional to the local
pulse strength. The ability to clearly identify the signal strength with an
atom's position degrades rapidly with the pulse width, however, particularly
considering the inherent noise associated with photon counting.

\section{2D Cluster}
\label{2dcluster}

\subsection{Scheme}

The full two-dimensional protocol works much the same way as the 1D case
discussed in Sec.~\ref{protocol}. In the simplest scheme, the beams are offset
horizontally as before, but now are centered vertically. The cluster must be
initially carved so that pulses measuring the physical qubits on one chain
have a negligible probability of projecting the states of physical qubits on
an adjacent chain. The horizontal chains representing computational qubits must
be sufficiently well-separated vertically. Na\"\i vely one might assume that
the chains would need to be separated by a distance at least greater than $na$,
with $n$ the offset defined in Eq.~(\ref{offset}). 

In practice the chains can be positioned much closer together, because of the
circular spatial `footprint' of the measuring pulse. Consider a beam that is
centered vertically on a given chain but is offset horizontally by $na$. If
the intensity on the first qubit is $I_0$, then the first qubits on chains
above and below will experience intensity $I_0e^{-2(ma)^2/r^2}$, where $ma$
is the separation between horizontal chains. If the desired probability of
projecting the first qubit on an adjacent chain is $p_m$, then the chain
separation is
\begin{equation}
m=\frac{r}{a}\sqrt{\frac{1}{2}\ln\left[
\frac{\ln(1-p_{\rm t})}{\ln(1-p_{\rm m})}\right]}.
\label{link}
\end{equation}
Thus the inter-chain separation grows more slowly (by a factor of $r/a$) than
the offset required to minimize inadvertent measurements on adjacent qubits on
the same chain, Eq.~(\ref{offset}). If for example one chooses
$p_{\rm t}=0.99$ and a tiny value $p_{\rm m}=10^{-10}$, then
$m\approx 3.5(r/a)$; with $r=4a$ and $r=10a$
one obtains $m\approx 14$ and $m\approx 35$, respectively. This separation is
adequate for all measurements on all chains (momentarily postponing issues
related to links between horizontal chains, discussed in detail below). One
simply requires that all the measured qubits in the various chains have the
same horizontal coordinate (column index).

Recall that prior to each measurement in the 1D protocol, one needs to perform
of order $2n$ rotations. These prepare the target qubit and those to its right
that might be inadvertently measured, and restore qubits previously measured
and soon-to-be-remeasured to the state $\ket{1}$. In 2D, the number of measured
qubits, and therefore the number of preparatory rotations, at each stage is of
order $\pi n^2$. This requires a solution to a $n^2\times n^2$ linear system
of equations whose cost scales as $n^4$, a significant classical overhead.
Note that the beam irradiating a given chain impinges on chains above and
below. So the necessity of performing preparatory rotations implies that
measurements of qubits on chains separated by distances smaller than $na$
cannot be done in parallel, unless the solution of the relevant linear
equations involves all rows of physical qubits. In practice, this
apparent lack of operational parallelism is no serious impediment: in the
worst-case scenario considered above with $r=10a$, there are only
$4.4$ chains per full offset.

The main extension of the 1D protocol to the 2D cluster is how to implement 
the teleportation primitive on the links between horizontal chains, in order
to simulate an entangling gate between computational qubits. As was discussed
in Sec.~\ref{1WQC}, no feed-forward is required when single-qubit measurements
teleport Clifford gates. Because the $CZ$ gate belongs to the Clifford group,
with suitable preparation of the relevant qubits (i.e.\ application of total
rotation of $\pi$) the inter-chain links (hitherto referred to as simply links)
should be unaffected by inadvertent measurements. That said, the outcomes of
measurements on links need to be unequivocally known, since they will affect
the feed-forward criteria for future measurements on chains.

Making well-defined measurements on qubits in vertically oriented links is
extremely difficult, however. Consider the situation where the target qubit
on a given chain is the first qubit of a vertical link. Using Eq.~(\ref{prob}),
the probability of measuring the first link qubit is close to $p_{\rm t}$.
Indeed, Eq.~(\ref{link}) states that $m$ qubits along the link from the target
will be measured with probability $p_{\rm m}$ or greater. If we wish that this
probability be at least smaller than the probability of inadvertently
measuring a qubit to the right of the target along the chain
$p_{\rm m'}<p_{\rm f}$, then the number of link qubits that will be measured
is $m'=\sqrt{2n+1}\approx m/2$ using Eq.~(\ref{offset}). Thus, the states of
individual link qubits will be impossible to uniquely determine.

The solution to this apparent conundrum is to form inter-chain links that are
essentially diagonal, with the `zig-zag' pattern shown in Fig.~\ref{zigzag}. 
The first two qubits in the link must be vertically oriented. Recall that the
cluster is formed by first initializing all cluster qubits to $\ket{+}$
followed by the $CZ$ gate, and one needs to avoid forming a box graph. With
this geometry, the maximum number of inadvertent measurements will be the
number of qubits in the column to the right of the target. It is clear from
the figure that this number will often be three: one on the chain and two
nearby in the link. More importantly, the maximum number of qubits intentionally
measured simultaneously will also be three. Even without the ability to spatially
resolve the signal emanating from three atoms, however, the individual states
of the three qubits can still be obtained with minimal operational overhead as
shown explicitly below.

\subsection{Protocol}

\begin{figure}[t]
\includegraphics[keepaspectratio,width=\columnwidth]{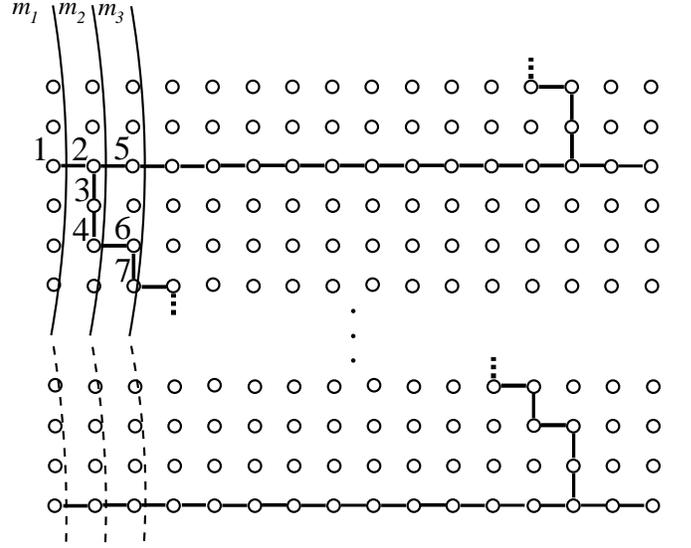}
\caption{The full 2D protocol. Initially, qubit 1 in the first column of the
upper chain is measured by a beam that is offset horizontally but centered
vertically (the right-most edge of the beam is labeled $m_1$). Immediately
thereafter, the first qubit on the second chain is measured with the same
horizontal offset, as is the first in the third chain, etc. Next, qubits 2, 3,
and 4 in the second column are simultaneously measured ($m_2$), as is the
second qubit in the lower chain, and so on. Measurements continue in this
fashion until the logical state is entangled between the two chains.}
\label{zigzag}
\end{figure}

The 2D protocol differs primarily in that simultaneous measurements of chain
and link qubits must be performed. Suppose the first target is in column 1, on
the upper chain just left of the first link qubit, as illustrated in 
Fig.~\ref{zigzag}. It is assumed to be in the
state $\ket{\psi}=\alpha\ket{0}+\beta\ket{1}$ with $\alpha$, $\beta$ complex
coefficients satisfying $|\alpha|^2+|\beta|^2=1$. If all entangled qubits are
first rotated by $\pi$, after measurement of qubit 1 the resulting
logical state is
\begin{eqnarray}
\rho_{1}&\approx&(1-p_{\rm f}-p_{\rm f}^2-p_{\rm f}^3)||X^{m_1}H\ket{\psi}||
\nonumber \\
&+&p_{\rm f}||X_1^{m_2}H_1H_2X_1^{m_1}H_1|\tilde{\psi}\rangle||\nonumber \\
&+&p_{\rm f}^2||X_2^{m_3}H_2X_1^{m_2}H_1H_2X_1^{m_1}H_1|\tilde{\psi}\rangle||
\nonumber \\
&+&p_{\rm f}^3||X_2^{m_4}H_2X_2^{m_3}H_2X_1^{m_2}H_1H_2X_1^{m_1}H_1
|\tilde{\psi}\rangle||\nonumber \\
&=&(1-p_{\rm f}-p_{\rm f}^2-p_{\rm f}^3)||X^{m_1}H\ket{\psi}||
\nonumber \\
&+&p_{\rm f}||X_1^{m_2}Z_1^{m_1}H_2|\tilde{\psi}\rangle||
+p_{\rm f}^2||X_2^{m_3}X_1^{m_2}Z_1^{m_1}|\tilde{\psi}\rangle||
\nonumber \\
&+&p_{\rm f}^3||X_2^{m_4}Z_2^{m_3}X_1^{m_2}Z_1^{m_1}H_2
|\tilde{\psi}\rangle||.
\label{2drho1}
\end{eqnarray}
This expression assumes that inadvertent measurements are possible only on
qubits in column 2 (labeled qubits 2, 3, and 4 in Fig.~\ref{zigzag}), in the
vicinity of the upper chain. The small differences
in the probabilities of inadvertently measuring qubits down the link are
neglected for clarity, and are all set to $p_{\rm f}$. Note that to define
the current logical state one can ignore measurements beyond the qubit that
connects to the original logical state. Also, in practice one can ignore terms
with prefactors higher than $p_{\rm f}^2$, since these are much smaller than
the probabilities of inadvertently measuring second-nearest neighbors which 
are ignored in the current treatment. In any case, the outcomes of these
inadvertent measurements will be determined shortly by intentional
measurements.

The most important new feature in the density matrix (\ref{2drho1}) is the
appearance of the state 
$|\tilde{\psi}\rangle\equiv\alpha\ket{00}+\beta\ket{11}$. When a
measurement is performed on the chain qubit that is also the first link
qubit, the chain and link qubits become inherently entangled. The logical
state must therefore be written in a two-particle basis with 
$\ket{0}\mapsto\ket{00}$ and $\ket{1}\mapsto\ket{11}$, with the first and
second elements representing the chain and link qubits, respectively. Of
course, this is how the link is able to simulate a $CZ$ gate between two
chains: upon measurement of the last link qubit, which is located on the
second chain, the two qubit basis corresponds to chain 1 and 2. The
subscripts on the single-particle operators in Eq.~(\ref{2drho1}) therefore
correspond to the qubit index in the two-particle basis.

After rotating all nearby unmeasured cluster qubits by $\pi$, the
measurement of atoms in the second column (qubits 2-4 in Fig.~\ref{zigzag})
can be performed. The outcome is
\begin{eqnarray}
\rho_{2}&\approx&(1-2p_{\rm f}-2p_{\rm f}^2)||X_2^{m_4}Z_2^{m_3}X_1^{m_2}
Z_1^{m_1}H_2|\tilde{\psi}\rangle||\nonumber \\
&+&p_{\rm f}||X_1^{m_5}X_2^{m_4}Z_2^{m_3}Z_1^{m_2}X_1^{m_1}H_2H_1|\tilde{\psi}
\rangle||\nonumber \\
&+&p_{\rm f}||X_2^{m_6}Z_2^{m_4}X_2^{m_3}X_1^{m_2}Z_1^{m_1}|\tilde{\psi}
\rangle||\nonumber \\
&+&p_{\rm f}^2||X_2^{m_7}Z_2^{m_6}X_2^{m_4}Z_2^{m_3}X_1^{m_2}Z_1^{m_1}H_2
|\tilde{\psi}\rangle||\nonumber \\
&+&p_{\rm f}^2||X_2^{m_6}X_1^{m_5}Z_2^{m_4}X_2^{m_3}Z_1^{m_2}X_1^{m_1}H_1
|\tilde{\psi}\rangle||,
\end{eqnarray}
ignoring terms of order $p_{\rm f}^3$ and higher. How is it possible to
uniquely determine the outcome $\ket{m_2m_3m_4}$? If no signal is obtained,
the three-qubit state is definitely $\ket{111}$. If a weak signal indicates
only one atom is in $\ket{0}$, then the possible outcomes are $\ket{011}$,
$\ket{101}$, and $\ket{110}$. Subsequent rotation of the qubit 2 by $\pi$ will
either yield $\ket{111}$ or two qubits in $\ket{0}$, which can be easily
verified by re-measurement of the three qubits. If the latter outcome is
obtained, then re-rotating qubits 2 and 3 both by $\pi$ will unequivocally
determine the correct initial output. The same approach works for an original
measurement indicating two $\ket{0}$ outcomes, and trivially for $\ket{000}$
or $\ket{111}$. Even if the relative
signal strengths corresponding to different numbers of $\ket{0}$ outcomes
cannot be distinguished, there are only seven different combinations that need
to be attempted (by repeated rotations and measurements) in order to convert
the output to $\ket{111}$ and thereby eliminate the fluorescence.

Measurements continue in this columnwise fashion until the link connects
with the lower chain. The signal strength from link qubits on measurements of
chain 1 qubits will gradually decrease, becoming undetectable soon after
the midpoint between chains. Before this column is reached, measurement of
chain 2 qubits will pick up signal from the link qubits, maintaining the
flow of information down the link. If the fluorescence noise is too great,
then one would simply move the center of the measuring beam vertically
between the two chains to improve the signal.

\section{Summary and Discussion}
\label{summary}

We have described in detail a protocol for deterministic quantum computation
in the one-way model, where multiple physical qubits are simultaneously
measured. This approach is well-suited to implementation with ultracold
atoms confined in optical lattices, where the measuring beam is much wider
than the separation between qubits. A central assumption of the approach is
that the center of the lasers can nevertheless be positioned with high
accuracy. Our strategy makes use of the Gaussian profile of lasers used in
these systems for performing both rotations and measurements. For rotations,
the Gaussian profile allows for a unique rotation on each cluster qubit to be
performed by superimposing rotations on all physical qubits. This enables
the direct generation of the desired computational cluster state, and to
teleport the desired unitaries during the one-way computation. For
measurements, the laser center is chosen to be far removed from the qubit of
interest in order to minimize the possibility of making inadvertent
measurements on nearby qubits. With additional overhead in terms of physical
qubits, however, the protocol can accommodate an arbitrary number of unknown
outcomes of measurements accidentally performed on nearby contiguous qubits.

Each measurement, which effects a gate on the computational qubits, is
performed in four steps. First, all qubits that will be irradiated by the
measuring beam, as well as several qubits beyond, must first have their phases
suitably prepared. This is accomplished by applying a rotation pulse centered
on each physical qubit. Second, the relevant cluster qubits are measured.
Third, if more than one qubit is intentionally measured, then these need to
be further rotated and subsequently re-measured. Fourth, if previous
measurement outcomes will incorrectly apply a future rotation due to an
inadvertent measurement, then this error needs to be fixed by a finite set of
future measurements.

Because teleportation errors caused by inadvertent measurements need to be
fixed by extending the number of measurements of qubits in the horizontal
chains, it is not known {\it a priori} how many chain qubits will be needed to
effect the desired unitary. This is somewhat problematic because the
computational cluster, i.e.\ the entire structure of horizontal chains and
vertical links, needs to be formed in advance of any measurements. The
simplest solution is to make all the inter-link distances sufficiently long
that the total probability of performing the correct unitary will be high. 
The probability of continued failure after $\ell$ attempted corrections is
$1/2^{\ell+1}$. If the desired probability of success is $1-\epsilon$ then the
number of correction attempts must be $\ell > \log_2(1/\epsilon)-1$. In order
to protect against inadvertent measurements of $m$ neighbors, an arbitrary
unitary will require a chain of $3(lm+1)$ qubits for even $m$ or $3[l(m+1)+1]$
for odd $m$. Once the correct unitary is accomplished, measuring the remaining
qubits will teleport Clifford gates with trivial consequences.

An attractive feature of this scheme is that in principle it could partially
mitigate the second most significant impediment in the implementation of 1WQC
with ultracold atoms in optical lattices: stray fluorescence from measurements
accidentally projecting distant qubits. Because photons are emitted in all 
directions, the probability of inadvertently measuring qubits a distance $d$
from the target decreases like $d^{-2}$. Simply buffering the chains with 32
extra qubits (each rotated by $\pi$ to teleport Clifford gates) reduces the
deleterious effects of stray light by a factor over 1000.

While the 1WQC scheme outlined in this manuscript can account for the
possibility of having inadvertently measured qubits, it is currently not
designed to protect against the small but finite possibility that the
{\it target} is not actually projected by the measurement. A central
assumption throughout has been that the laser coupling, measurement time, and
the decay rate are all sufficiently large that an intentional measurement is
performed with high certainty. The issue might seem irrelevant because any
given target is consistently re-measured as the protocol moves down a chain
or inter-chain link, so it will quickly become projected if it was not
initially, and any error in assumptions will be made manifest. The problem is
that rotating a qubit that is believed to have been already projected, but is
in fact entangled with yet-unmeasured cluster qubits, will possibly teleport
an unknown Pauli byproduct gate on subsequent measurements. This issue is
however beyond the scope of the current calculations, and will be the focus of
future work.

\acknowledgments
The authors are grateful to Mark Tame and Terry Rudolph for stimulating
discussions in the early stages of the work, and to Michael Garrett for
insightful comments. This research was supported by the Natural Sciences
and Engineering Research Council of Canada and the Canada Foundation for
Innovation.

\end{document}